\documentclass[aps,prx,groupedaddress,twocolumn, longbibliography, superscriptaddress]{revtex4-1}

\usepackage{graphicx}
\usepackage{amssymb, amsmath}
\usepackage{bm}

\newcommand{\diff}{\mathrm{d}}

\newcommand{\BES}{\mathcal{S}}
\newcommand{\n}[1]{\langle \hat n_{#1} \rangle}
\newcommand{\nn}[2]{\langle \hat n_{#1} \hat n_{#2} \rangle}
\newcommand{\N}{\langle \hat N \rangle}
\newcommand{\nN}[1]{\langle \hat N \hat n_{#1}  \rangle}

\newcommand{\bn}{{\boldsymbol{n}}}
\newcommand{\no}{\hat{n}}

\begin{document}

\title{Pump-power-driven mode switching in a microcavity device\\ and its relation to
Bose-Einstein condensation}
\author{H.~A.~M.~Leymann}
\email{ham.leymann@gmail.com}
\affiliation{Institut f{\"u}r Theoretische Physik, Otto-von-Guericke-Universit{\"a}t
  Magdeburg, Postfach 4120, D-39016 Magdeburg, Germany}
\affiliation{Max-Planck-Institut f{\"u}r Physik komplexer Systeme, N{\"o}thnitzer Strasse 38, D-01187 Dresden, Germany}

\author{D.~Vorberg}
\email{dv@pks.mpg.de}
\affiliation{Max-Planck-Institut f{\"u}r Physik komplexer Systeme, N{\"o}thnitzer Strasse 38, D-01187 Dresden, Germany}

\author{T.~Lettau}
\affiliation{Institut f{\"u}r Theoretische Physik, Otto-von-Guericke-Universit{\"a}t
  Magdeburg, Postfach 4120, D-39016 Magdeburg, Germany}

\author{C.~Hopfmann}
\affiliation{Institut f\"{u}r Festk\"{o}rperphysik,
  Technische Universit\"at Berlin, Hardenbergstra{\ss}e 36, D-10623
  Berlin, Germany}

\author{C.~Schneider}
\affiliation{Technische Physik and Wilhelm Conrad R\"ontgen Research
  Center for Complex Material Systems, Physikalisches Institut,
  Universit\"at W\"urzburg, Am Hubland, D-97074 W\"urzburg, Germany}

\author{M.~Kamp}
\affiliation{Technische Physik and Wilhelm Conrad R\"ontgen Research
  Center for Complex Material Systems, Physikalisches Institut,
  Universit\"at W\"urzburg, Am Hubland, D-97074 W\"urzburg, Germany}

\author{S.~H\"ofling}
\affiliation{Technische Physik and Wilhelm Conrad R\"ontgen Research
  Center for Complex Material Systems, Physikalisches Institut,
  Universit\"at W\"urzburg, Am Hubland, D-97074 W\"urzburg, Germany}
\affiliation{SUPA, School of Physics and Astronomy, University of St Andrews, St Andrews, KY16 9SS, United Kingdom}

\author{R.~Ketzmerick}
\affiliation{Max-Planck-Institut f{\"u}r Physik komplexer Systeme, N{\"o}thnitzer Strasse 38, D-01187 Dresden, Germany}
\affiliation{Technische Universit{\"a}t Dresden, Institut f{\"u}r Theoretische Physik and Center for Dynamics, D-01062 Dresden, Germany}

\author{J.~Wiersig}
\affiliation{Institut f{\"u}r Theoretische Physik, Otto-von-Guericke-Universit{\"a}t
  Magdeburg, Postfach 4120, D-39016 Magdeburg, Germany}

\author{S.~Reitzenstein}
\affiliation{Institut f\"{u}r Festk\"{o}rperphysik,
  Technische Universit\"at Berlin, Hardenbergstra{\ss}e 36, D-10623
  Berlin, Germany}

\author{A.~Eckardt}
\email{eckardt@pks.mpg.de}
\affiliation{Max-Planck-Institut f{\"u}r Physik komplexer Systeme, N{\"o}thnitzer Strasse 38, D-01187 Dresden, Germany}




\begin{abstract}
We investigate the switching of the coherent emission mode of a bimodal microcavity device, occurring when the pump power is varied. We compare
experimental data to theoretical results and identify the underlying mechanism
to be based on the competition between the effective gain on the one hand and
the intermode kinetics on the other. When the pumping is ramped up, above a threshold the mode with the largest effective gain starts to emit coherent light, corresponding to lasing. In contrast, in the limit of strong pumping it is the intermode kinetics that determines which mode acquires a large occupation and shows coherent emission. We point out that this latter mechanism is akin to the
equilibrium Bose-Einstein condensation of massive bosons. Thus, the mode
switching in our microcavity device can be viewed as a minimal instance of
Bose-Einstein condensation of photons.
We, moreover, show that the switching from one cavity mode to the other occurs always via an intermediate phase where both modes are emitting coherent light and that it is
associated with both superthermal intensity fluctuations and strong
anticorrelations between both modes.
\end{abstract}



\maketitle

\section{Introduction}
\label{sec:introduction}


The development of optical cavities \cite{he_whispering_2013,
vahala_optical_2003, kryzhanovskaya_whispering-gallery_2014, cao_dielectric_2015}
has led to (laser-)devices with an almost vanishing lasing
threshold \cite{lermer_high_2013,nomura2010}. Bimodal (micro)lasers have been realized in various types
of systems, such as ring lasers \cite{lett_macroscopic_1981, mandel_optical_1995},
vertical-cavity surface-emitting lasers \cite{sondermann_two-frequency_2003}, its quantum-dot micropillar variant
\cite{leymann_intensity_2013, leymann_strong_2013}, and 2D photonic crystal
cavity lasers \cite{zhukovsky_switchable_2007}. In these systems, the switching
of the mode showing coherent emission \cite{sondermann_two-frequency_2003, choquette1994temperature,
sun1995polarization, martin1997polarization,rubio_coexistence_2001, ackemann2001characteristics,
marconi_asymmetric_2016} has gathered substantial interest due to potential
technical applications as optical flip-flop memories, tunable sensitive switches
\cite{ge_interaction-induced_2016, alharthi_circular_2015} and as a simple
realization of non-equilibrium phase transitions~\cite{agarwal_higher-order_1982,
gartner_laser_2016}.

In this article, we study the switching of the coherent emission mode in bimodal micropillar
lasers occurring when the pump power is ramped up. By comparing experimental data
to theoretical results based on a phenomenological model, we identify the basic mechanism underlying the mode
switching to be the competition between effective gain on the one hand and the
intermode kinetics on the other. Namely, the mechanism that selects which of the modes shows coherent emission is found to be fundamentally different for weak pumping (just above the threshold) and in the limit of strong pumping. For weak pumping, the \emph{selected} mode (i.e.~the mode selected for coherent emission) is characterized by the largest effective gain and coherent emission corresponds to lasing. In contrast, for strong pumping, the selected mode depends neither on the coupling to the gain medium nor on the loss rates of both modes. Instead it is determined completely by the intermode kinetics, i.e.\ by processes that transfer photons from one mode to the other. We show that this mechanism is formally identical to the one leading to
Bose-Einstein condensation of an ideal gas of massive bosons (i.e.~with a conserved particle number) in contact with a
thermal bath. Therefore, the mode switching in our system
can be viewed as a minimal instance of Bose-Einstein
condensation of photons.

The question, whether a system of photons (or bosonic quasiparticles) with
non-conserved particle number can undergo Bose-Einstein condensation in a
similar way as a thermal gas of massive bosons has raised considerable interest
in the last decade. Here the problem to be overcome is to achieve a quasi-equilibrium
situation, where a single mode acquires a macroscopic occupation via a thermalizing
kinetics, despite the non-equilibrium nature of the system resulting from
particle loss to be balanced by pumping. Beautiful experiments, showing that
such a situation can indeed be achieved, have been conducted in systems of
exciton-polaritons \cite{KasRicKun2006,balili_bose-einstein_2007, DenHuaYam2010,wertz_spontaneous_2010, CarCiu2013, byrnes_exciton-polariton_2014,fischer_spatial_2014}
magnons \cite{DemDemDzyEtAl2006, BunVol2008, VaiAhoJarEtAl2015, FanOlfWuEtAl2016}, and photons
in dye-filled cavities \cite{KlaSchVew2010}.
While the microcavity device investigated in this article is simpler than these
systems, in the sense that it is described in terms of two relevant modes only,
it captures one of the most important aspects of photon condensation in a minimal
fashion, namely that the condensate mode is selected not by pumping but rather by
the kinetics of the photons.

\begin{figure}
  \includegraphics[width=0.7\linewidth]{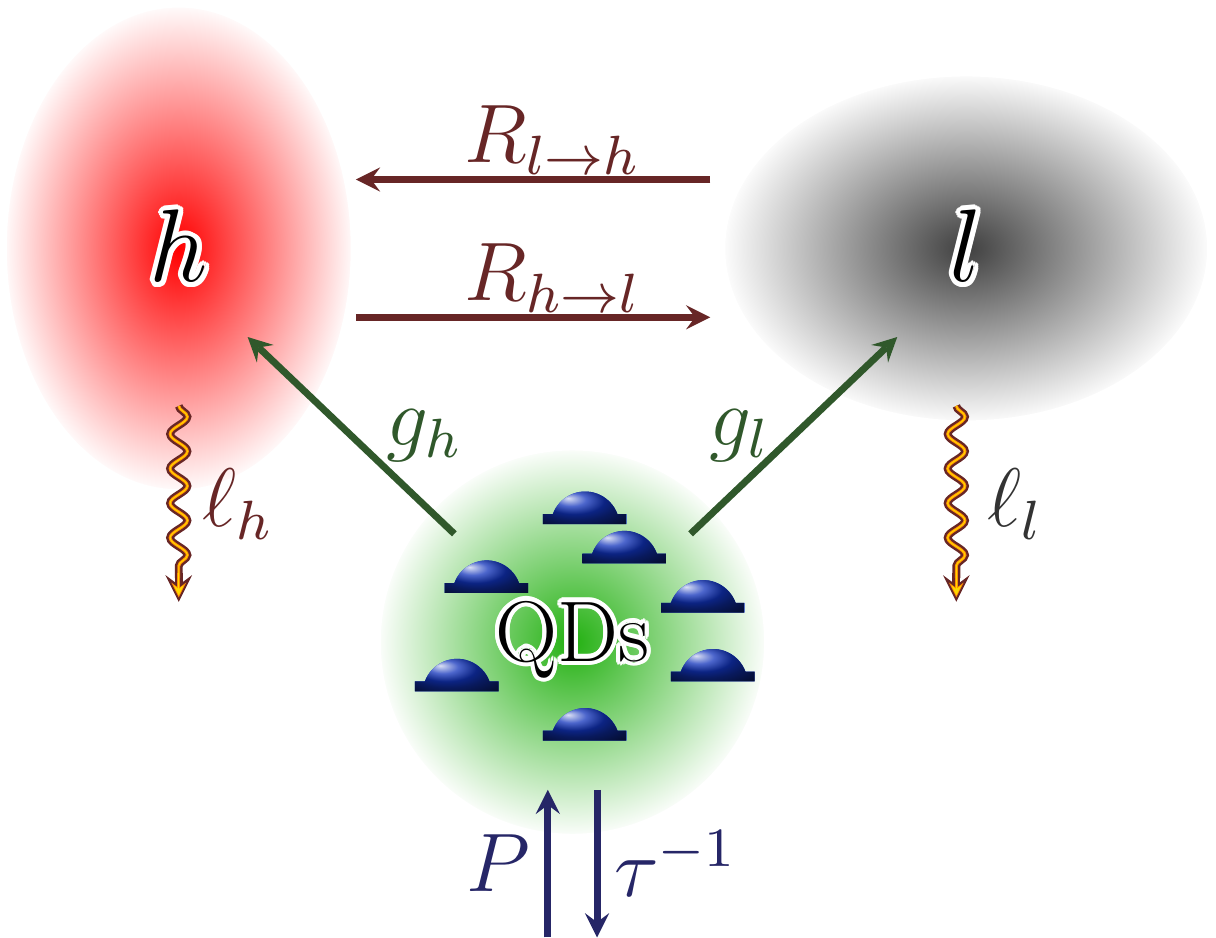}
  \caption{\label{fig:modelillustration}Illustration of the relevant processes in our bimodal system. The intermode kinetics between the modes $h$   and $l$ is determined by the rates $R_{i\to j}$. The modes lose their photons with rates $\ell_i$. This loss is compensated by new excitation from the quantum dots (QDs) with gain rates $g_i$. The quantum dots are pumped with rate $P$ and lose their excitations spontaneously with rate $\tau^{-1}$. }
\end{figure}

\begin{figure}
  \includegraphics[width=0.95\linewidth]{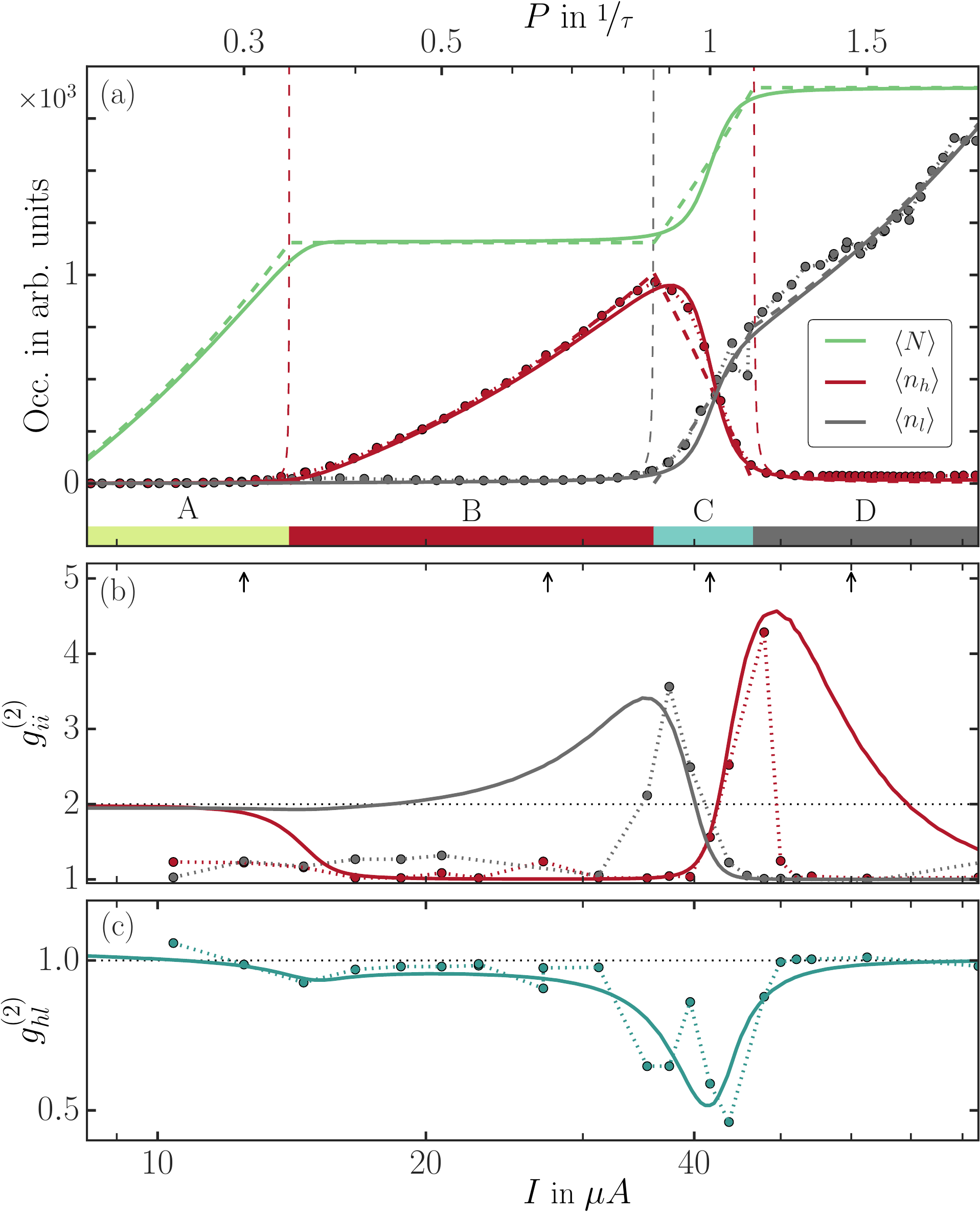}
  \caption{\label{fig:exp_vs_theory} Measured and calculated microcavity input-output
    characteristics (high-effective-gain mode in red,
    low-effective-gain mode in gray). The injection current $I$ is proportional to
    the pump rate $P$. Here, circles (connected to guide the eye), solid curves, and
    dashed curves refer to experimentally measured data, exact Monte-Carlo results,
    and asymptotic mean-field theory, respectively. Panel (a) depicts the mean
    occupations of the modes and the number of excited carriers (light green) in
    arbitrary units. The colored bars at the bottom of this panel mark the phases
    determined by the asymptotic theory (see Sec.~\ref{sec:appl_asymp}).
    Panel (b) shows the autocorrelation $g_{ii}^{(2)}$ for both modes in
    comparison with the value 2 expected for thermal emission (dotted black line).
    Panel (c) depicts the crosscorrelation $g_{hl}^{(2)}$ in comparison to the
    value 1 for correlated emission (dotted black line). For the theoretical curves,
    we use $s=0$, $G=0.77$ as well as (in units of the spontaneous loss rate
    $\tau^{-1}$) $g_{h}=1.6\cdot 10^{-3}$, $g_{l}=2.1\cdot 10^{-3}$,
    $\ell_{h} = 2.2\cdot 10^{-2} $, $\ell_{l} = 3.8\cdot 10^{-2}$,
    $R_{l\to h}=1.7\cdot 10^{-4}$, $A_{h\to l}=8.5\cdot 10^{-6}$
    (see App.~\ref{app:extracting}).}
\end{figure}

The starting point of our theoretical decription of a bimodal
micropillar is a generic phenomenological master equation describing the relevant
processes of the system \cite{rice_photon_1994, leymann_intensity_2013}:
the coupling between the pumped medium and the cavity modes, loss, and the
intermode kinetics (Fig.~\ref{fig:modelillustration}). Such birth-death models have also been used to study (analogs of Bose-Einstein condensation in) population dynamics \cite{KneKruWeb2013, KneWebKru2015}, transport \cite{PhysRevLett.103.090602, evans2005nonequilibrium}, and networks \cite{PhysRevLett.86.5632} as well as quantum gases of massive bosons \cite{vorberg_generalized_2013, vorberg_nonequilibrium_2015}.
This approach, which is different from the microscopic modeling pursued in former studies
of bimodal (micro) lasers \cite{Haken1963,tehrani_coherence_1978, mandel_optical_1995,
leymann_intensity_2013, khanbekyan_unconventional_2015,
ge_interaction-induced_2016,redlich_mode_2016, fanaei_effect_2016} and interacting
exciton-polariton systems, e.g., Refs.~\cite{Krizhanovskii2008, Dagvadorj2015, Altman2015}, provides
excellent agreement with the experimental data (Fig.~\ref{fig:exp_vs_theory}).
In order to treat these equations analytically, we work out an approximation scheme that combines the theory of Bose selection, which was recently developed to describe (non-equilibrium) Bose condensation of  ideal gases of massive bosons with conserved particle number \cite{vorberg_generalized_2013}, with particle loss and the coupling to a pumped reservoir (gain medium). We justify this approximation by exact numerical simulations. Somewhat counter-intuitively, our theory shows that it is the limit of strong pumping, where the selected mode is determined by the intermode kinetics, which is described by terms that are formally identical to those appearing in the description of massive bosons and their equilibrium condensation.
We also find that the mode switching occurs via an intermediate phase where
both modes are emitting coherently (see phase diagram in Fig.~\ref{fig:phase} below).

We investigate also the statistical properties of our system. In bimodal lasers,
where one mode dominates the emission for all pump rates, the non-lasing mode
exhibits usually super-thermal intensity fluctuations
\cite{lett_macroscopic_1981, leymann_intensity_2013} and
the emission of both modes is strongly anticorrelated \cite{lett_investigation_1986, mandel_optical_1995,roy_first-passage-time_1980,
murthy_monte_1985,redlich_mode_2016}.
In the situation where mode switching occurs, we find that the super-thermal
intensity fluctuations of the non-selected mode and strong anti-correlations occur
whenever a mode starts or ceases to be selected. We show that these experimentally observed
statistical properties can be described theoretically by an effective reduction
of the spontaneous inter-mode transitions caused by mode interactions.

This paper is organized as follows. In Sec.~\ref{sec:experiment} the experimental
setup is presented. The theoretical description in terms of a master equation is
introduced in Sec.~\ref{sec:Master Equation}. An analytical theory of the mode
switching and its relation to Bose-Einstein condensation is then worked out in
Sec.~\ref{sec:Mean-Field Theory}. Finally, in Sec.~\ref{sec:Phtostat}, we
investigate the statistical properties of the system, before coming to the
conclusions in Sec.~\ref{sec:conclusion}.

\section{Experiment}
\label{sec:experiment}

Electrically-pumped quantum-dot micropillars are fabricated by etching of
a planar AlAs/GaAs distributed Bragg reflector $\lambda$-cavity in which a single
active layer of self-assembled In$_{0.3}$Ga$_{0.7}$As quantum dots is embedded
centrally \cite{kistner_demonstration_2008}.
A detailed description can be found in Ref.~\cite{Boeckler2008}. The micropillar
used in this study has a diameter of $3.0~\mu m$. Due to the strong confinement
of light, the micropillars exhibit a spectrum of discrete modes. The fundamental
modes $HE_{1,1}$ are composed of two orthogonally linearly polarized components,
which are ideally energetically degenerate. In reality, however, asymmetries in
the manufacturing process, which results in slightly elliptical structures, lift
the energetical degeneracy of the fundamental modes \cite{he_whispering_2013,
reitzenstein_alasgaas_2007}. The resulting mode splitting of the micropillar
used in this study is $(42 \pm 2)~\mu eV$. Besides a finite mode splitting
the two fundamental modes also exhibit slightly different quality factors,
$14000 \pm 1500$ and $12500 \pm 1500$, respectively. The different spectral and local
overlaps of the modes with the gain medium and the modes polarization alters their coupling to the quantum dot emitters. As a
consequence, the former mode (mode $h$) is characterized by a \emph{higher}
effective gain [i.e.\ gain-loss ratio, see Eq.~\eqref{eq:effective_gain_modes}
below] than the latter one, which has a \emph{lower} effective gain (mode $l$).

A high-resolution ($25~\mu eV$) micro-electroluminescence setup is used to characterize the micropillars spectrally at cryogenic temperatures of $15\,K$.
A linear polarizer as well as a $\lambda/4$-plate in front of the monochromator enables polarization-resolved spectroscopy. For statistical analysis via the autocorrelation function $g^{(2)}_{ii}(\tau=0)$ with zero delay time of the emission a fiber-coupled Hanbury Brown and Twiss setup is used.
The temporal resolution of the Hanbury Brown and Twiss configuration based on fast Si avalanche photon diodes is $\tau_{\mathrm{res}}= 40~ps$.
For measuring the equal-time crosscorrelation function $g^{(2)}_{hl}(\tau=0)$
between the orthogonally polarized micropillar modes the emission is selected by a polarization maintaining 50/50~beamsplitter and the split beam is directed to a second identical spectrometer - with the polarizer in front of the second monochromator oriented orthogonally to the first.
These equal-time correlations are defined by
\begin{align}\label{eq:g2}
  g^{(2)}_{ij}(\tau=0)=\frac{\langle \hat b^\dagger_i \hat b^\dagger_j \hat b_i \hat b_j\rangle}{\langle \hat b^\dagger_i \hat  b_i \rangle\langle \hat b^\dagger_j \hat b_j \rangle}\equiv g^{(2)}_{ij},
\end{align}
where $\hat b_{i}$ is the bosonic annihilator operator of mode $i$.
The emission characteristics of the CW-pumped micropillar are shown in
Fig.~\ref{fig:exp_vs_theory}. In (a) the output characteristics of modes $h$ (red)
and $l$ (gray) are shown as a function of the injection current $I$ (quantifying
the pumping strength). Up to about $13~\mu A$ (phase A) both modes are below
threshold and show a small increase in output emission only. Between $13~\mu A$
and $36~\mu A$ (phase B) the mode $h$ is selected and its emission increases
strongly while that of mode $l$ remains small. Between $36~\mu A$ and
$50~\mu A$ (phase C) both modes are selected, but while the emission intensity of
mode $l$ increases that of mode $h$ decreases. Beyond $50~\mu A$ (phase D)
only mode $l$ is selected.
In Fig. \ref{fig:exp_vs_theory} (b) and (c) the zero-delay autocorrelation
$g_{ii}^{(2)}$ and crosscorrelation $g_{hl}^{(2)}$ functions are
plotted. Note that, due to finite temporal resolution of the avalanche photodiode
detectors and the low coherence time of the modes, we could not properly
resolve $g_{ii}^{(2)}$ experimentally for values above $1$ in phase A and B
\cite{ulrich_photon_2007}.

\section{Master Equation}
\label{sec:Master Equation}

Our starting point for the theoretical description of the system is a
phenomenological
master equation for the probabilities $\rho^{\bn}_N$ to find the system in a state
with $N$ excited emitters and photon numbers $\bn=(n_{h},n_{l})$ in the
high- and the low-effective-gain mode. It takes the form
\begin{align}
  \frac{\diff}{\diff t}\rho^{\bn}_N = C_\text{laser}(\rho)+C_\text{kin}(\rho)
  \label{eq:rate_eq}
\end{align}
and shall be solved for the steady-state state obeying
\begin{align*}
  \frac{\diff}{\diff t}\rho^{\bn}_N=0.
\end{align*}

The first term on the right hand side of the master equation (\ref{eq:rate_eq})
describes how photons leave and enter the cavity modes via loss and coupling to
the emitters. It is given by
\begin{align}
C_\text{laser}(\rho)
    =&P\left[\rho^{\bn}_{N-1}-\rho^{\bn}_N\right]
    -\tau^{-1}[N\rho^{\bn}_N-(N+1)\rho^{\bn}_{N+1}]
 \nonumber \\ &
    -\sum_i g_{i}[N(n_{i}+1)\rho^{\bn}_N
                -(N+1)n_{i}\rho^{{\bn}-{\bf e_i}}_{N+1}]
 \nonumber\\&
    -\sum_i \ell_{i}[n_{i}\rho^{\bn}_N
                -(n_{i}+1)\rho^{{\bn}+{\bf e_i}}_N],
\end{align}
where $P$ and $\tau^{-1}$ denote the pump and the loss rate of the emitters,
respectively, $g_i$ quantifies the gain of cavity mode $i$ from the emitters,
and $\ell_i$ is the loss rate of cavity mode $i$. An additional or removed
photon in mode $i$ is denoted by $\pm{\bf e_i}$,
[i.e.,~${\bm e}_h=(1,0)$, ${\bm e}_l=(0,1)$]. The modes $h$ and $l$ are defined
by the higher and lower effective gain $g_i/l_i$, respectively, so that we obtain
an effective-gain ratio
\begin{align}
  \label{eq:effective_gain_modes}
  G \equiv \frac{g_l/\ell_l}{g_h/\ell_h}<1.
\end{align}
The terms contained in $C_\text{laser}(\rho)$ are sufficient for a theoretical
description of single-mode lasing in mode $h$ \cite{rice_photon_1994}.

The second term of the master equation (\ref{eq:rate_eq}) captures the intermode
kinetics and reads
\begin{align}\label{eq:Ikin}
C_\text{kin}(\rho)=&-\sum_{i,j}R_{i\to j}\big[n_{i}
    (n_{j}+s)\rho^{\bn}_N
    \nonumber\\&
      -(n_{i}+1)(n_{j}-1+s)\rho^{{\bn}+{\bf e_i}-{\bf e_j}}_{N}\big].
\end{align}
It is characterized by the rates $R_{i\to j}$ for a transition from mode $i$
to mode $j$. The rate asymmetry of the direct intermode transitions
\begin{align}
  A_{i\to j}=R_{i\to j}-R_{j\to i}
\end{align}
is generally nonzero.
The origin of this rate asymmetry was attributed to stimulated scattering
due to carrier population oscillations, e.g., in coupled
photonic crystal nanolasers \cite{marconi_asymmetric_2016}.
Furthermore
asymmetric backscattering of electromagnetic waves was observed in optical
 microcavities \cite{WKH08, Wiersig11, POL16}.
In our system the rate asymmetry $A_{h\to l}$ is positive,
\begin{equation}
  A_{h\to l} > 0,
\end{equation}
so that the low-effective-gain mode $l$ is favored by the intermode kinetics.
The parameter $s$ in Eq.~\eqref{eq:Ikin} quantifies the ratio between
spontaneous and induced intermode transitions. Its natural value is
$s = 1$. However, a reduction of $s$ to values $s<1$ is a simple
way to effectively capture inter-mode interactions that lead to a relative
enhancement of transitions into strongly occupied modes. We will show below that, while $s$ has (practically) no impact on the phase transition and the
mean occupation(s) of the selected mode(s), it does affect the occupation of the
non-selected mode. Only for $s<1$, the master equation can describe the
experimentally observed super-thermal fluctuations $g^{(2)}_{il}>2$ of the
non-selected mode. The numerical data shown in Figs.~\ref{fig:exp_vs_theory},
\ref{fig:photon_statistic}, and \ref{fig:photon_statisticpmeps} are obtained for
$s=0$.

The form of the term $C_\text{kin}(\rho)$ capturing the intermode kinetics is
identical to that of the master equation for an ideal gas of massive bosons in
contact with an environment \cite{vorberg_generalized_2013}. If such a system of
massive bosons is coupled to a thermal environment characterized by the
temperature $T$, the intermode rates obey
$R_{j\to i}/R_{i\to j} =\exp[-\Delta_{ij}/(k_BT)]$ with Boltzmann constant $k_B$
and energy splitting $\Delta_{ij}=E_i-E_j$ between modes (single-particle states)
with energy $E_i$. In the quantum degenerate regime of low temperature or high
boson density, the system will form a Bose-Einstein condensate in the
single-particle state of lowest energy. When increasing the total number of bosons
$N_B$ in this regime, the occupation of an excited mode $i$ approaches the finite
value $\langle\no_i\rangle=(e^{E_i/k_BT}-1)^{-1}$, while the ground-state occupation
increases linearly with $N_B$. Even for a finite number of discrete energy levels
$i$, in the limit $N_B\to\infty$ this behavior clearly describes Bose-Einstein
condensation as the macroscopic occupation of one single-particle state (see
e.g.\ Ref.~\cite{vorberg_nonequilibrium_2015}).

The most intriguing result of this paper, shown below, is that the behavior
of the bimodal system in the limit of strong pumping strength $P$ closely resembles
that of a Bose-Einstein condensed gas of massive bosons in equilibrium. Remarkably
here the selected mode does not depend on the effective gain, but is determined
exclusively by the intermode transitions $R_{i\to j}$.
This countner intuitive result is related to the fact that the intermode kinetics
scales quadratically, but gain and loss only linearly with the mode occupations.

It is instructive to define the dimensionless parameter $\varepsilon$,
\begin{equation}\label{eq:DB}
  \frac{R_{l\to h}}{R_{h\to l}} \equiv \exp(-\varepsilon),
\end{equation}
which can be interpreted as the ratio
$\varepsilon=\Delta^\text{eff}_{hl}/(k_B T_\text{eff})$ of an effective energy
splitting $\Delta^\text{eff}_{hl}$ between both modes and
an effective temperature $T_\text{eff}$. In the limit of strong pumping, the
intermode kinetics makes the photons condense into the mode corresponding to the
lower effective energy. That is for $\varepsilon>0$ or $A_{h\to l}>0$
($\varepsilon<0$ or $A_{h\to l}<0$) a
Bose-Einstein condensate of photons is formed in mode $l$ (mode $h$). In contrast,
it is always the mode $h$, characterized by the higher effective gain, that
starts lasing when the pump power $P$ is ramped up. Thus, a rate
asymmetry $A_{h\to l}>0$ implies a switching from lasing in mode $h$ to condensation in
mode $l$, when the pump power is ramped up. In the following section, we derive
an analytical theory, which describes this effect and establishes the analogy
to equilibrium Bose-Einstein condensation for strong pumping.

\section{Kinetic Theory}
\label{sec:Mean-Field Theory}

\subsection{Mean-field approximation}
\label{sec:Rateequ}

In order to obtain a closed set of kinetic equations for the mean mode occupations
$\n{i}=\sum_{\bn,N}\rho_N^{\bn} n_{i}$ and the mean number of excited emitters
$\N=\sum_{{\bn},N}\rho_N^{\bn} N$, we perform the mean-field approximation
\begin{equation}
  \nn{i}{j}\approx\n{i}\n{j} \quad\text{and} \quad \nN{i}\approx\N\n{i}.
\label{eq:mf}
\end{equation}
This approximation, which ignores non-trivial two-particle correlations, is later
justified by comparing it to exact solutions of the full master equation
(\ref{eq:rate_eq}) obtained from Monte-Carlo simulations (see
Fig.~\ref{fig:exp_vs_theory}). Employing it, we derive kinetic equations of
motion for the mean occupations:
\begin{align}
  \frac{\diff}{\diff t} \n{i} =& \sum_j \big[R_{j\to i}\n{j}(\n{i}+s)-
  R_{i\to j}\n{i}(\n{j}+s)\big]
  \label{eq:mf_n}\nonumber\\&
  + g_{i} \N\big( \n{i} + 1\big) - \ell_{i} \n{i},
  \\
  \frac{\diff}{\diff t} \N =& -\frac{\N}{\tau } + P - \sum_i g_{i} \N\big( \n{i} +1 \big).
  \label{eq:mf_N}
\end{align}

\subsection{Asymptotic theory}
\label{sec:Asmptotic_theory}

In a next step, for the sake of finding an analytical expression for the mean
occupation(s) of the selected mode(s), in Eq.~\eqref{eq:mf_n} we neglect
spontaneous processes relative to corresponding stimulated ones,
$(\langle\hat{n}_i\rangle+a)\simeq \langle\hat{n}_i\rangle$ with $a=1,s$.
This approximation is valid asymptotically in the limit of large occupations of
the emitters and the selected mode(s). Note, for high-$\beta$
cavities, where almost the entire spontaneous emission goes into the cavity modes, this
assumption is plagued by the strong presence of spontaneous emission at the
first threshold. Still, this assumption is valid in the coherent emission
regime and, in particular, when describing transitions, where the selected modes
change. Using the asymptotic approximation above, the stationary solution of
Eq.~\eqref{eq:mf_n} obeys
\begin{align}
  \n{i}\Big[g_{i}\N-\ell_{i}+\sum_{j} A_{j\to i}\n{j}\Big]=0.
  \label{eq:naive_guess}
\end{align}
This equation is solved by mean occupations that can be divided into two classes.
For the non-selected modes $i\notin\BES$, one finds the trivial
solution $\n{i}=0$, whereas the occupations of the selected modes
$i\in\BES$ obey a linear set of equations:
\begin{align}
  \sum_{j\in\BES} A_{j\to i}\n{j}&=\ell_{i}-g_{i}\N,\quad\forall i\in\BES.
\label{eq:n_sel}
\end{align}
The occupation of the non-selected states, which vanishes in leading order
[Eq.~\eqref{eq:naive_guess}], can be computed in the next order of our approximation.
For this purpose, we take into account those terms, which are linear in the number of excited emitters $\N$ or the occupations $\n{i}$ of the selected modes. We find:
\begin{align}
  \n{i}=\frac{g_{i}\N + s\sum_{j\in\BES} R_{j\to i}\n{j}}{\ell_{i}-g_{i}\N-\sum_{j\in\BES}A_{j\to i}\n{j}} \quad \forall i\notin\BES.
\label{eq:n_non_sel}
\end{align}
The dependence of the number of excited emitters on the pumping can be obtained by analogue reasoning,
\begin{align}
  \N =\frac{P}{\tau^{-1} +  \sum_{j\in \BES}g_{j}\n{j}}.
  \label{eq:pumping2}
\end{align}

Following the strategy recently employed for massive bosons
\cite{vorberg_generalized_2013}, the set of selected modes $\BES$ can now be
determined by the physical requirement that all modes (both selected and
non-selected modes) must have positive occupations,
\begin{align}
  \n{i}\ge0 \ \forall i.
\end{align}
For the non-selected modes this implies that the denominator of
Eq.~\eqref{eq:n_non_sel} has to be positive. Thus, the set of selected modes and
their occupations are independent of the parameter $s$ since $s$ occurs
neither in Eq.~\eqref{eq:n_sel} nor in the denominator of
Eq.~\eqref{eq:n_non_sel}.

\subsection{Phase diagram of the bimodal microcavity system}
\label{sec:appl_asymp}

\begin{figure}
  \centering
  \includegraphics[width=1\linewidth]{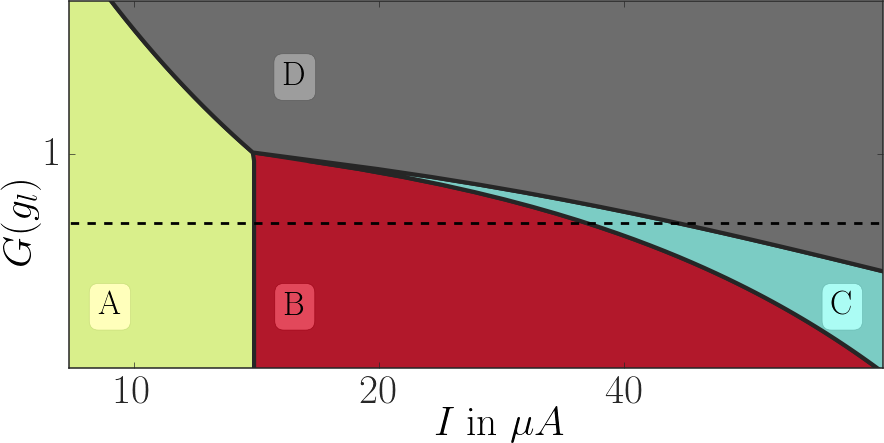}
  \caption{\label{fig:phase}
    Phase diagram with phases (A) no selected mode, (B) mode $h$ is selected, (C)
    both modes are selected, and (D) mode $l$ is selected. Plotted versus the
    injection current $I$ that is proportional to pump rate $P$ and the ratio of the effective gain
    rates $G=g_{l}\ell_{h}/g_{h}\ell_{l}$ with fixed
    $g_{h}\ell_{l}/\ell_{h}$.
    Except for $g_l$ all parameters are the same as in Fig.~\ref{fig:exp_vs_theory}.
    The dashed line indicates the value of $g_l$ used in
    Fig.~\ref{fig:exp_vs_theory}. While for $G<1$
    all phases are present when increasing the pump rate $P$, for
    $G>1$ the mode $l$, that is favored by the
    direct mode coupling becomes actually the high-effective-gain mode so that no
    mode switching occurs.}
\end{figure}

Based on the asymptotic theory, we can now compute the non-equilibrium phase
diagram. The concept of selected modes, which appeared
naturally in the asymptotic theory, clearly separates the parameter space into
different phases, where no, one, or both modes are selected. A transition, where
one of the non-selected modes becomes selected and starts emitting coherent light, is indicated
by the divergence of its occupation described by Eq.~\eqref{eq:n_non_sel}. In
turn, a selected mode ceases to be selected when its occupation, obtained by solving
Eq.~\eqref{eq:n_sel}, drops to zero. In this section we will use this argument to
compute the phase boundaries analytically. The resulting phase diagram is
depicted in Fig.~\ref{fig:phase}.

In phase A, for small pumping power $P$, neither mode is selected, $\BES=\{\}$.
According to Eq.~\eqref{eq:n_non_sel} the mode occupations read
\begin{align}
  \n{i}_A\approx\frac{1}{\ell_{i}/(g_{i}\N)-1}\ ,
\end{align}
and the number of excited emitters increases linearly with the pump rate,
\begin{equation}
  \N_A=\tau P .
\end{equation}
When the pumping is increased and reaches
\begin{align}
  P_{AB}\approx\frac{\ell_h}{g_h\tau},
\end{align}
the occupation of the high-effective-gain mode diverges indicating the transition
to a regime where this mode is selected and starts emitting coherently
(see Eq.~\eqref{eq:n_non_sel}). Since before the transition no mode is selected yet,
this asymptotic estimate for the critical pumping strength cannot be expected to
mark precisely the threshold in the high-$\beta$ limit.
Note, however, that the estimates for further thresholds at $P_{BC}$ and $P_{CD}$
are not plagued by this problem, since they occur in the regime where at least one mode is selected already.

In phase B the high-effective-gain mode is selected, $\BES=\{h\}$, and the number of excited emitters is clamped at the threshold value
[see Eq.~\eqref{eq:n_sel} and Fig.~\ref{fig:exp_vs_theory} (a)]
\begin{align}
  \N_{B}=\frac{\ell_{h}}{g_{h}}.
\end{align}
The excitation provided by increasing the pumping is directly transferred to the
selected mode~\cite{siegman_lasers_1986}. Consequently, the occupation of the selected
high-effective-gain mode depends linearly on the pump rate
[use Eq.~\eqref{eq:pumping2}]
\begin{align}
  \n{h}_B = \frac{P}{\ell_{h}} - \frac{1}{\tau g_{h}}.
\label{eq:n1B}
\end{align}
The occupation in the non-selected low-effective-gain mode is given by [see Eq.~\eqref{eq:n_non_sel}]
\begin{align}
  \n{l}_B = \frac{g_{l}\N_B +s R_{h\to l}\n{h}_B}{\ell_{l}-g_{l}\N_B-A_{h\to l}\n{h}_B}\label{eq:n2_B}.
\end{align}
In a case where the mode-coupling rates would favor the high-effective-gain mode
$(A_{h\to l}<0)$, Eq.~\eqref{eq:n2_B} would be valid for all pumping powers
$P>P_{AB}$. However, in our case, where the mode-coupling rates favor the
low-effective-gain mode $(A_{h\to l}>0)$, increasing the pump rate
(and with it also $\n{h}_B$), eventually leads to the divergence of the right-hand
side of Eq.~(\ref{eq:n2_B}). This occurs at the pump rate
\begin{align}
  P_{BC} = \frac{\ell_{h}}{g_{h}}\left[\frac{1}{\tau } + \frac{1}{A_{h\to l}}(g_{h}\ell_{l}-g_{l}\ell_{h})\right]
\end{align}
and indicates the transition to phase C.

In phase C both modes are selected, $\BES=\{h,l\}$. The
number of excited emitters increases again linearly with the pump
rate $P$,
\begin{align}
  \N_C=P\left[\frac{1}{\tau } + \frac{1}{A_{h\to l}}(g_{h}\ell_{l}-g_{l}\ell_{h})\right]^{-1}.
\label{eq:pumpingC}
\end{align}
The occupations of the high- and low-effective-gain mode de- and increase linearly with $\N_C$, respectively,
\begin{align}
  \n{h}_C &= -\frac{g_{l}\N_C-\ell_{l}}{A_{h\to l}}\label{eq:n1C},
  \\
  \n{l}_C &= \frac{g_{h}\N_C-\ell_{h}}{A_{h\to l}}\label{eq:n2C}.
\end{align}
When the number of the emitters $\N_C$ reaches $\ell_{l}/g_{l}$, the occupation
$\n{h}_C$ becomes zero, indicating the transition to the phase where this mode is no
longer selected. The threshold pump rate $P_{CD}$ can be obtained from
Eq.~\eqref{eq:pumpingC} analogously to the expression for $P_{BC}$ and reads %
\begin{align}
  P_{CD} = \frac{P_{BC}}{G}.
  \label{eq:P_CD}
\end{align}
Thus the extent of phase C is determined by the inverse effective gain
ratio $G$ [Eq.~\eqref{eq:effective_gain_modes}].

In phase D, only mode $l$ is selected, $\BES=\{l\}$. The number of excited emitters remains at the higher threshold value
\begin{align}\label{eq:ND}
  \N_D=\frac{\ell_{l}}{g_{l}},
\end{align}
and the occupation of the selected mode $l$ reads
\begin{align}
  \n{l}_D = \frac{P}{\ell_{l}} - \frac{1}{\tau g_{l}}.
  \label{eq:n2D}
\end{align}
The occupation of the non-selected mode $h$ is given by
\begin{align}
  \n{h}_D = \frac{g_{h}\N_D+sR_{l\to h}\n{l}_D}{\ell_{h}-g_{h}\N_D+A_{h\to l}\n{l}_D}\label{eq:n1_D}.
\end{align}
The crucial difference to Eq.~\eqref{eq:n2_B} is that for $A_{h\to l}>0$ an
increase of $\n{l}_D$ cannot produce a further root in the denominator of
Eq.~\eqref{eq:n1_D}. Thus no further transition will occur [unless other parameters change as well when the pump power is ramped up].

Figure~\ref{fig:phase} shows the phase diagram resulting from the asymptotic
theory with respect to the effective-gain ratio $G$ (varied by varying the
gain rate $g_l$) and the pumping strength $P$ (proportional to the injection current $I$).
While the precise shape of the phase boundaries depends on the parameters (and
which of them are varied in order to modify the effective-gain ratio $G$), the
topology of the phase diagram is generic.
For $G<1$ (and $A_{h\to l}>0$), the system always undergoes a sequence of three
transitions between the phases A, B, C, D when the pump rate is increased.
For $G>1$ (and $A_{h\to l}>0$), where the mode labeled $l$ actually becomes
the high effective gain mode, only a single transition from phase A to
phase D occurs.
In summary, when the pump power $P$ is ramped up, the system starts lasing in the
mode characterized by the higher effective gain, whereas in the limit of strong
pumping the selected mode is the one favored by the intermode kinetics. Moreover,
the switching from selection of mode $h$ to selection of mode $l$ has to occur via an
intermediate phase, where both modes are selected (unless the system is
fine-tuned to $G=1$).

The data plotted in Fig.~\ref{fig:exp_vs_theory} corresponds to a cut through
this phase diagram following the dashed horizontal line in Fig.~\ref{fig:phase}.
The different phases obtained from the asymptotic theory are indicated by the
colors at the bottom of panel (a). In Fig.~\ref{fig:exp_vs_theory}(a), we can
clearly see that the mean occupations obtained from the asymptotic theory
(dashed lines) nicely reproduce the exact solution (solid line) of the master
equation (\ref{eq:rate_eq}), which was obtained by Monte-Carlo simulations (for
a detailed description of the method see
Ref.~\cite{vorberg_nonequilibrium_2015}). This agreement justifies both the
mean-field approximation and the asymptotic theory. More importantly, the theoretical curves also
describe the experimental data (circles in Fig.~\ref{fig:exp_vs_theory}) very well.
In appendix~\ref{app:extracting} we explain how the parameters of our
model are determined.

\subsection{Relation to Bose-Einstein condensation}
We have seen that in the limit of strong pumping strength $P$ the selected mode
is determined exclusively by the intermode kinetics, which is described by the
rates $R_{i\to j}$. It is remarkable that in this limit neither the loss rates
of the modes $\ell_i$ nor their coupling $g_i$ to the emitters influences the
selection of the mode. This counter-intuitive result is related to the fact that
gain and loss scale only linearly with the mode occupations, whereas the rates for
the intermode kinetics have a quadratic dependence on the mode occupations.
It implies that the mechanism leading to a macroscopic (or large) occupation of
one of the modes is the same as the one that leads to the Bose-Einstein
condensation of massive bosons in contact with a thermal environment.
This is based on Bose-enhanced inter-mode kinetics (scattering) described, e.g., in
Ref.~\cite{vorberg_generalized_2013} on the basis of a rate equation comprising the same terms as $C_\text{kin}$ [Eq.~(\ref{eq:Ikin})].

Even though the system consists of two levels only, the notion of Bose
condensation becomes sharp in the limit $P\to\infty$, where the occupation of
mode $l$, given by Eq.~(\ref{eq:n2D}), approaches infinity, while that of mode
$h$, given by Eq.~(\ref{eq:n1_D}), remains below a finite value.
Note that also the number of excited emitters, given by Eq.~(\ref{eq:ND}),
remains at a finite value in this limit. This has the important consequence that
the occupation of the non-condensed mode $h$ is determined completely by the
intermode kinetics described by the rates $R_{i\to j}$. In the limit $P\to\infty$
it approaches
\begin{align}
  \n{h} \simeq \frac{s}{\exp(\varepsilon)-1},
\end{align}
where $\varepsilon$ is the parameter defined in Eq.~(\ref{eq:DB}). Thus,
for $s=1$ the occupation of mode $l$ precisely corresponds to that of an excited
state of energy $\varepsilon k_B T$ in a Bose condensed system of massive bosons
at temperature $T$. The fact that the ``excited-state'' occupation
$\langle \hat{n}_h\rangle$ (the depletion) approaches a constant value for strong
pumping, so that increasing the number of photons (bosons) in the system will only
increase the condensate occupation, is another clear analogy to equilibrium Bose
condensation. Irrespective of the value of $s$, for strong pumping the selection
of the coherent emission mode is a result of the intermode kinetics.

\section{Photon statistics}
\label{sec:Phtostat}

\begin{figure*}
\centering
  \includegraphics[width=1\linewidth]{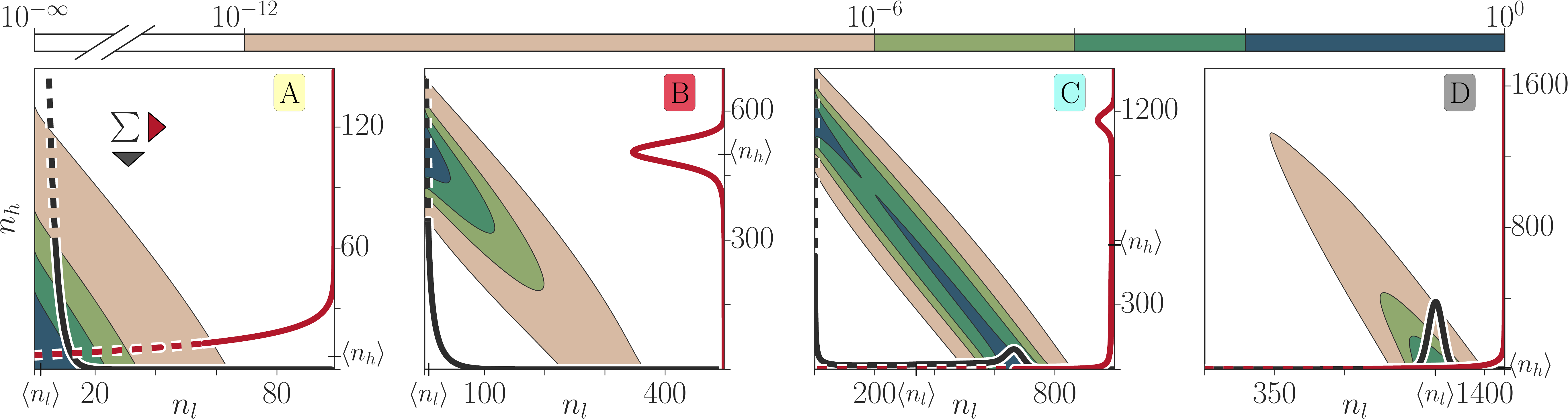}
  \caption{Two-mode photon statistics obtained from the reduced master equation (see Appendix \ref{app:reduced}) with single-mode statistics for mode $l$ at the bottom (black) and mode $h$ at the right (red) axis, for injection currents $I  = 12.5\,\mu A, 27.4\,\mu A, 41.7\,\mu A, 60\,\mu A$ (phase A, B, C, D). These pump rates are also indicated by arrows in Fig.~\ref{fig:exp_vs_theory}(b). The labels $\n{h}$ and $\n{l}$ mark the position of the expectation value for each mode. For a better comparison the single-mode statistics are depicted up to $0.025$\% in solid lines, while the remaining part is dashed. Note that plotted range of photon numbers increases from left to right.}
  \label{fig:photon_statistic}
\end{figure*}
In microlasers, where almost the entire spontaneous emission goes into a single mode, no sharp
intensity jump is visible at the lasing threshold \cite{rice_photon_1994,lermer_high_2013}. Instead the transition to coherent emission is indicated in the autocorrelation \cite{glauber_quantum_1963} of the emitted photons
\cite{strauf2006, ulrich_photon_2007, wiersig_direct_2009,
asmann_higher-order_2009, leymann_expectation_2014, chow_emission_2014,
jin_photon-number_1994}. To confirm the coherence properties of the emitted photons, we examine the photon statistics.
The equal-time photon correlation functions (\ref{eq:g2}), which we can write
like
\begin{align}
 g^{(2)}_{ij}=
 \begin{cases}
   \big(\nn{i}{i}-\n{i}\big)/\n{i}^2&\mbox{ if }i=j\\
   \nn{i}{j}/\n{i}\n{j}&\mbox{ if }i\neq j~,
  \end{cases}
\end{align}
measure the occupation number fluctuations of each mode and the
crosscorrelation between the modes, respectively.
In Fig.~\ref{fig:exp_vs_theory} (b) and (c) experimental and theoretical results for
$g^{(2)}_{ii}$ and $g^{(2)}_{hl}$ are depicted. The theoretical values
for $g^{(2)}_{ij}$ are determined by a numerically exact Monte-Carlo
simulation of the full master equation \eqref{eq:rate_eq}. As discussed above a change
from $g^{(2)}_{hh}=2$ to $g^{(2)}_{hh}=1$ indicates the first threshold where the coherent emission in mode $h$ sets in, as can be seen clearly in Fig.~\ref{fig:exp_vs_theory}.
For even stronger pumping, when entering or leaving phase $C$, in which both modes are selected, we observe pronounced anticorrelations between both
modes ($g_{hl}<1$) as well as superthermal intensity flucutations ($g_{ii}>2$) of
that mode $i$ that changes its state from non-selected to selected or vice versa.
Note that for our system parameters, phase C appears in a narrow interval of pump powers only so that its properties are overshadowed by those of the transitions BC and CD. As a result, the two minima of the crosscorrelations occurring at the transitions have merged to a single one.

In order to reproduce the measured superthermal intensity fluctuations at the transitions BC and CD theoretically, we have to choose $s<1$, corresponding to the presence of
intermode interactions. The best results are obtained for the value $s=0$, which
we used also in the simulations (the role of $s$ will be discussed in more detail
below and in Appendix~\ref{app:factorization}).

We will now investigate the signatures of the mode switching in the reduced
two-mode photon distribution
$\rho^{n_{h},n_{l}}=\sum_N\rho^{n_{h},n_{l}}_N$, which gives the
probability to find the system in a state with $n_{h}$ and
$n_{l}$ photons in mode $h$ and mode $l$, respectively. We compute this
quantity by solving either the full master equation (phases A and B) or the
reduced master equation for $\rho^{n_{h},n_{l}}$ (phases C and D, see
Appendix \ref{app:reduced} for details). Results for four different pump powers
$P$, corresponding to the four phases A to D, are depicted in
Fig.~\ref{fig:photon_statistic}. The corresponding single-mode distributions
$\rho^{n_{h}}=\sum_{n_{l}}\rho^{n_{h},n_{l}}$ and $\rho^{n_{l}}$
are shown as well.

\begin{figure}
\centering
  \includegraphics[width=1\linewidth]{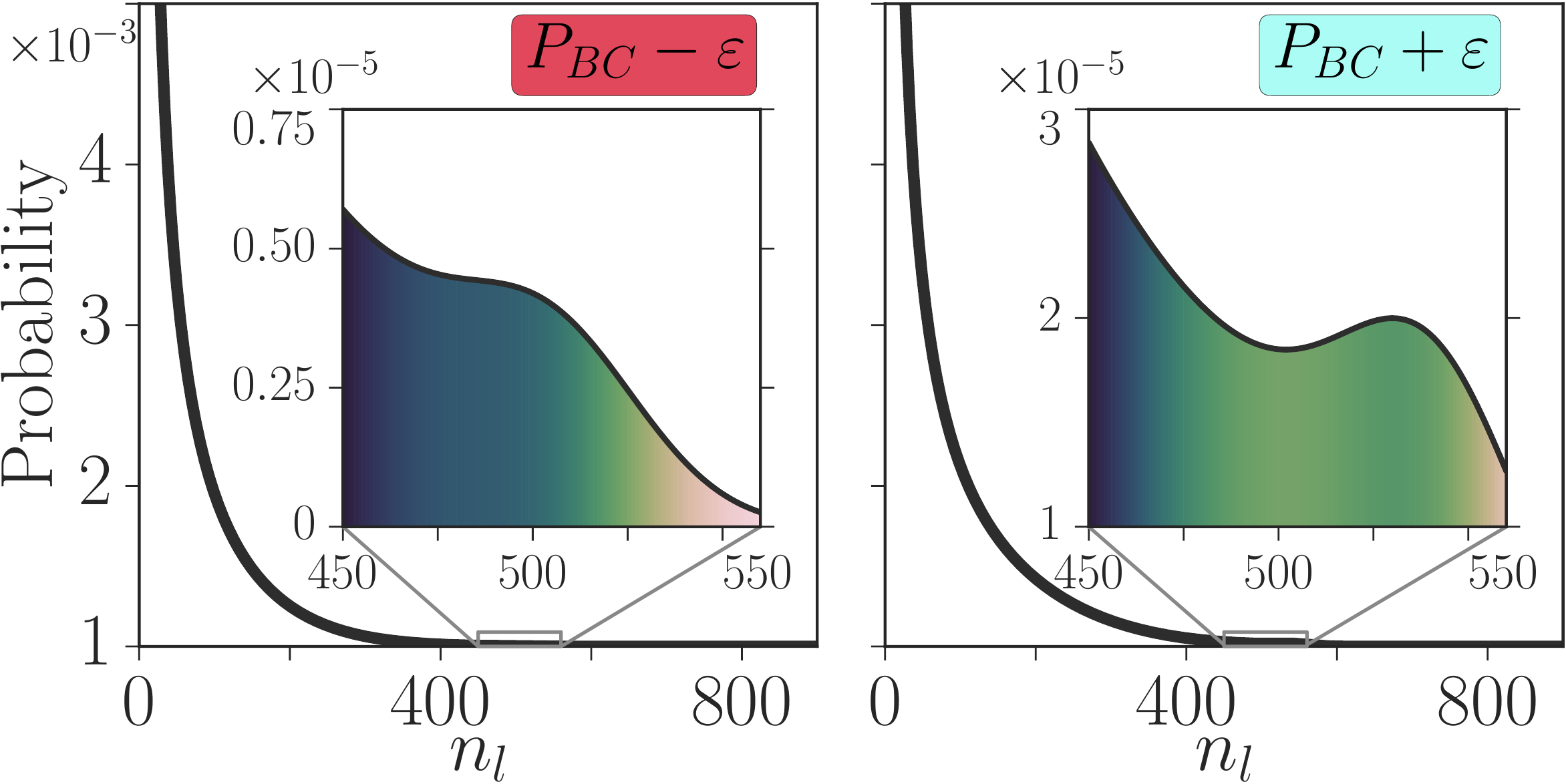}
  \caption{Photon statistics of mode $l$ ($\rho^{n_{l}}$) for pump rates below ($P_{BC}-\varepsilon$, left panel) and above ($P_{BC}+\varepsilon$, right panel) with $\varepsilon=0.016\mu A$. Only in the right panel the photon statistics exhibits a second local maximum. The system parameters are the same as in Fig.~\ref{fig:exp_vs_theory}.}
  \label{fig:photon_statisticpmeps}
\end{figure}

In the non-selected phase A the distribution possesses a single maximum at
${\bm n} = (n_h,n_l)=(0,0)$. The selection of mode $h$ (phases B and C) is
associated with a maximum of the distribution at
${\bm n}=(n^\text{max}_h,0)$ lying on the vertical axis, whereas the
selection of mode $l$ (phases C and D) with a maximum at
${\bm n}=(0,n^\text{max}_l)$ lying on the horizontal axis. Thus, in
phase $C$ where both modes are selected, the distribution possesses two local maxima
that are separated by a saddle point \footnote{This double peak structure can be
associated to mode switching in the time domain
\cite{lett_investigation_1986,redlich_mode_2016}.}.
The emergence of the second maximum when entering phase $C$, is visible also in
the occupation distribution of the mode that starts emitting coherently, as can be seen in
Fig.~\ref{fig:photon_statisticpmeps} showing $\rho^{n_l}$ directly before and
after the transition from $B$ to $C$. The build-up and later the presence of the
second maximum in $\rho^{n_l}$ is accompanied by the strong (superthermal) number
fluctuations in this mode.

\begin{figure}
  \centering
  \includegraphics[width=1\linewidth]{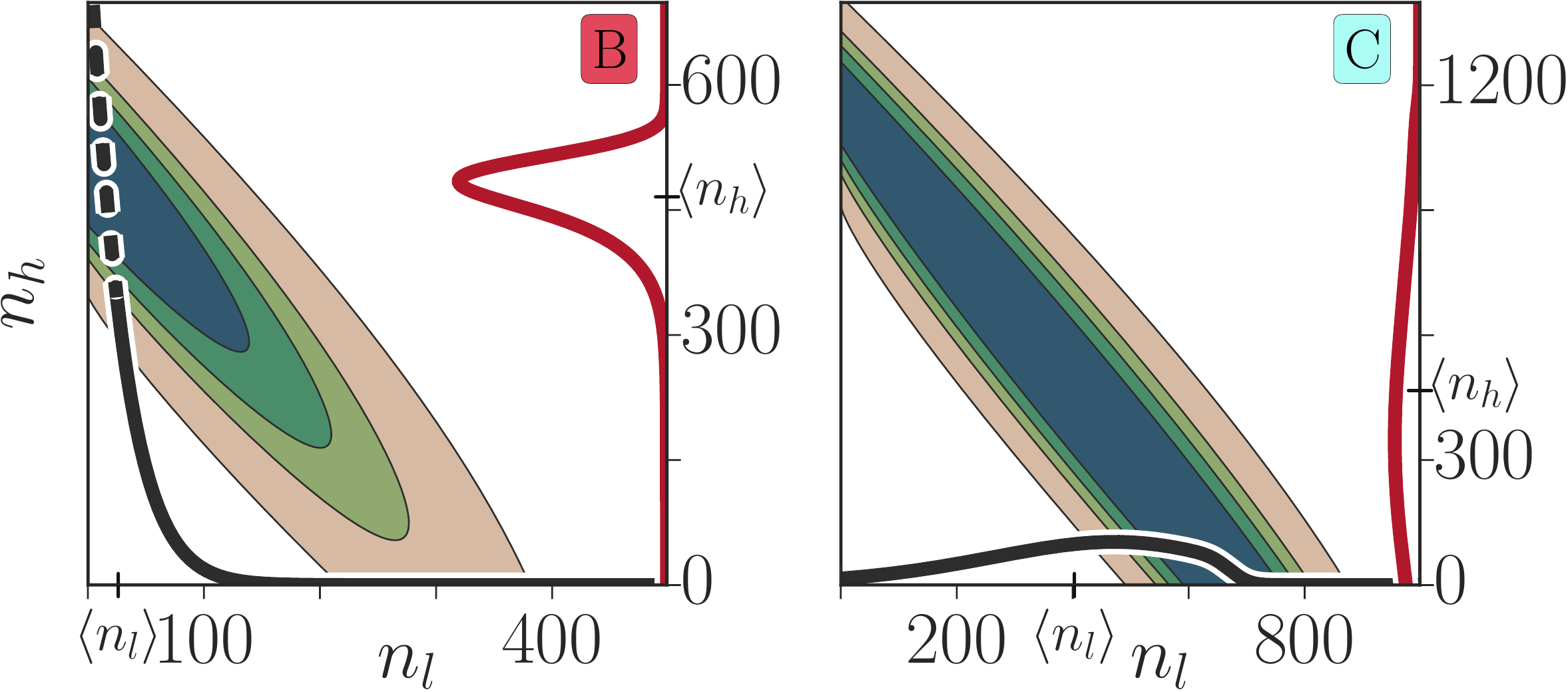}
  \caption{Two-mode photon statistics obtained from the reduced master equation for the same parameters as in Fig.~\ref{fig:photon_statistic} B and C but with fully included spontaneous transitions between the modes. In contrast to the case with suppressed spontaneous transitions, the statistics are no longer attracted to the axis. The colorbar is the same as in Fig.~\ref{fig:photon_statistic}. For a better comparison the single-mode statistics are depicted up to $0.015$\% in solid lines, while the remaining part is dashed.}
  \label{fig:photon_statistic_s}
\end{figure}

The impact of the effective parameter $s$ is illustrated in
Fig.~\ref{fig:photon_statistic_s}, where the two-mode distribution is shown for
$s=1$ in phases $B$ and $C$ for the same parameters used in the corresponding
panels in Fig.~\ref{fig:photon_statistic}. For $s=1$ states with zero occupation
in one of the modes (situated along the axes of the plot) are much less attractive
than for the value $s=0$ considered before. As a striking consequence, the selection
of both modes in phase C is not associated with two maxima in the distribution
anymore, but rather with a single central maximum. The transition from phase B to
phase C now corresponds to the shifting of the single maximum away from the
horizontal axis. As a result, for $s=1$ the system does not show the
experimentally observed superthermal photon number fluctuations, as we discuss
in more detail in Appendix ~\ref{app:factorization}. It is an interesting
observation that neglecting or taking into account the spontaneous emission between the modes has
such a strong impact on the statistical properties of the modeled bimodal system.
Note, however, that despite this strong impact on the statistics, the mean
occupations of the modes and the critical parameters for the phase
transition do not show a strong dependence on $s$. This is also the result
predicted by the asymptotic theory presented in the previous section.

\section{Conclusion}
\label{sec:conclusion}

In this paper we investigated the pump-power driven mode switching in a bimodal
microcavity. We presented experimental results and their explanation in
terms of a transparent analytical theory. 
In particular we found that the transition has to occur via an intermediate phase
where both modes are selected. Our theoretical
description reveals, moreover, a close connection to the physics of equilibrium
Bose-Einstein condensation in quantum gases of massive bosons. The mode switching
can, therefore, be viewed as a minimal instance of Bose-Einstein condensation of
photons and its demarcation to lasing.

We also investigated the statistical properties of the system and pointed out
that the mode switching is accompanied by superthermal intensity fluctuations as
well as anticorrelations between both modes. This observation can be technically
relevant, since a device producing a drastically increased occurrence rate of photon
pairs (with a very narrow linewidth \cite{khanbekyan_unconventional_2015}) could be
used to enhance two-photon excitation process used in fluorescence microscopy
\cite{jechow_enhanced_2013}. Furthermore a device that changes its predominantly
emitting mode in dependence of the injection current could be applied for optical
memories and other types of mode
management~\cite{zhukovsky_switchable_2007,lv_mode_2014}.

\begin{acknowledgments}
The authors thank A.~Musia{\l{}} for very helpful comments on the manuscript.
TL, DV, and HAML contributed equally to this work.
DV is grateful for support from the Studienstiftung des Deutschen Volkes. We acknowlege funding from the European Research Council under the European Union's Seventh Framework ERC Grant Agreeement No. 615613 and from the German Research Foundation (DFG) via Project No. Re2974/3-1 and the Research Unit FOR2414.  
\end{acknowledgments}

\appendix

\section{Extracting system parameters from the measured data}
\label{app:extracting}
The asymptotic theory describes the generic form of the mode switching and its
analytic expressions can thus be used to obtain the parameters of the master
equation model.
However, the theoretical parameters cannot be related directly to experimental
parameters due to the unknown proportionality factor $a_i$ between the intensity
of the emitted light $J_i$ and the occupation of modes, $\n{i}=a_i J_i$,
and the unknown excitation efficiency $b$ of the pumping with respect to the
injection current, $P=bI$.

The main properties of the switching are captured by the effective-gain ratio
$G$ which determines whether a switching occurs and the extent of phase C,
$\frac{P_{CD}}{P_{BC}}=G^{-1}$ [see Eq.~\eqref{eq:P_CD}]. This ratio can be
obtained in the following way:
Apply first a linear fit for the intensities of the selected modes $i$ in each
of the phases B, C, D,
\begin{align}
  J_i(I)|_R = A_{i R} I + B_{i R}.
\end{align}
The ratio $\frac{P_{CD}}{P_{BC}}=\frac{I_{CD}}{I_{BC}}$ is then determined
either by the  intersections of $J_h|_B(I_{BC})=J_h|_C(I_{BC})$ and
$J_l|_C(I_{CD})=J_l|_D(I_{CD})$ or
via the points where the occupation of the modes approach zero
$J_h|_C(I_{CD})=0$ and $J_l|_C(I_{BC})=0$.
Both procedures give similar values for the effective-gain ratio $G$ via
Eq.~\eqref{eq:P_CD}, namely 1.22 and 1.27, respectively. Thus determining $G$
does not require the knowledge of the excitation efficiency $b$ and the absolute
number of cavity photons via $a_i$.

The parameters $A_{h\to l}$, $g_h$, $g_l$, $\ell_h$, $\ell_l$, are extracted for comparison between theory and experiment
via the least squares method for all experimental data with $I< 80~\mu A$ and
are listed in the caption of Fig.~\ref{fig:exp_vs_theory}.
Since the time scale in Eq.~\eqref{eq:rate_eq} does not affect steady state properties, all parameters are measured in units of $\tau$.
The individual rates $R_{l\to h}$ and $R_{h\to l}$ do not affect the
asymptotic theory, only the rate asymmetry $A_{l\to h}$ does [as discussed above].
But the correlation function $g^{(2)}_{ll}$ does depend on the individual rates,
so that $R_{l\to h}$ is chosen to reproduce this correlation function.

In the experiment the orientation of a polarization filter is chosen parallel to
the passive cavity modes at the inversion point. Due to the interactions induced
by the common gain medium the polarization-resolved spectrum exhibits a double
peak structure, indicating that each polarization direction contains small
portions of the other mode \cite{khanbekyan_unconventional_2015}.
To make numerical and experimental fluctuations comparable we take into account that
in each polarization a small fraction of the other mode is detected by introducing
the mixing
\begin{align}
 n^{\mathrm{meas}}_{h,l}=(1-c)\,n_{h,l}\pm c\, n_{l,h},
 \label{eq:mixing}
\end{align}
where $n^{\mathrm{meas}}_{h,l}$ denotes the quantity that is measured.
The mode mixing prevents the experimental observed fluctuations of the non-selected
mode from increasing monotonically in phase D [see Fig.~\ref{fig:exp_vs_theory}(b)].
Comparing the theoretical results with the experimental data, we find the optimal
value of the mixing parameter to be very small, $c=8 \cdot 10^{-3}$. Its impact of
the mixing is negligible in phases A, B, and C. It is taken into account in the
numerical data presented in Figs.~\ref{fig:exp_vs_theory} and
\ref{fig:exp_vs_theory_s}, where the index `$\mathrm{meas}$' is dropped.

\section{Reduced density matrix}

\begin{figure}
 \includegraphics[width=0.95\linewidth]{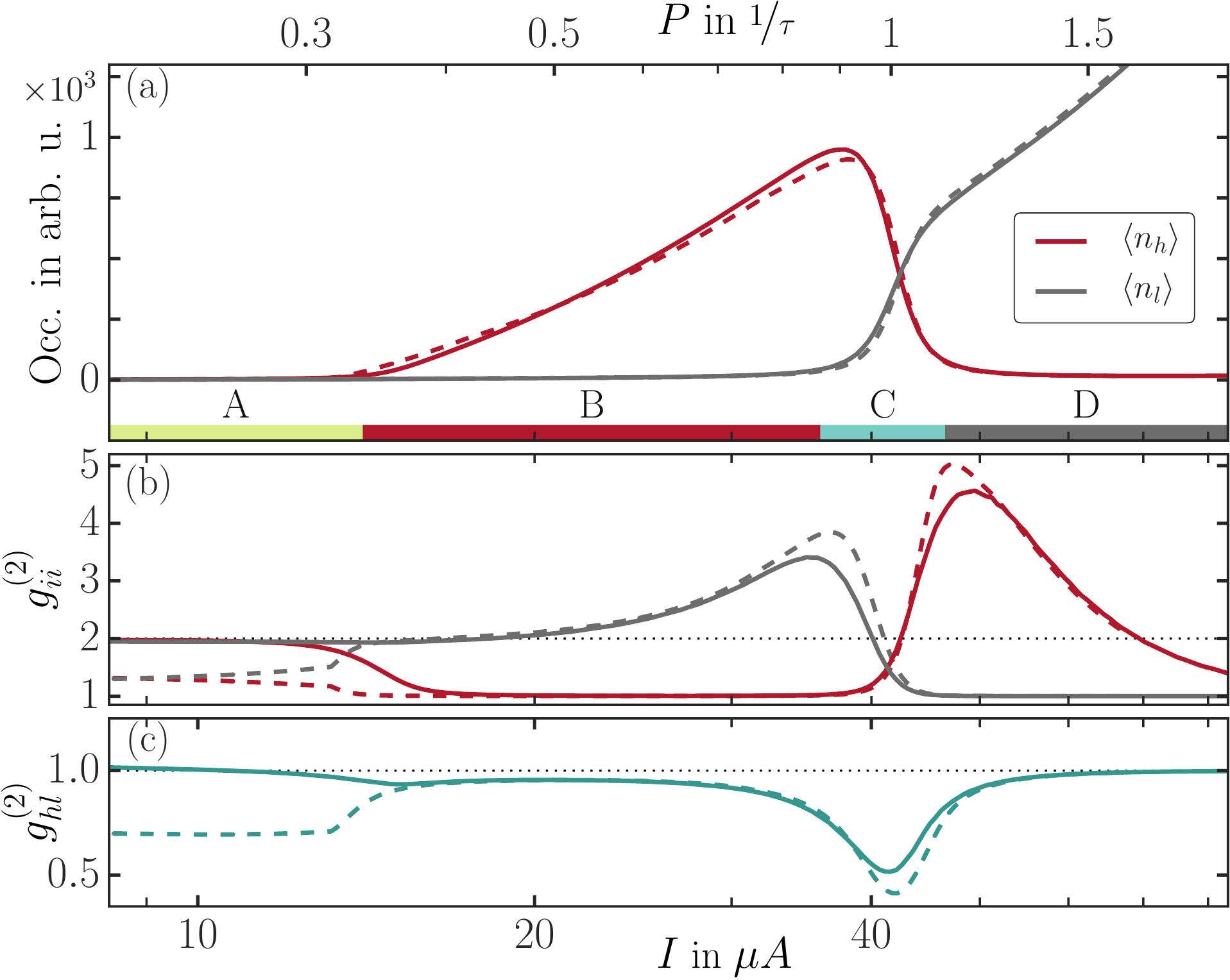}
 \caption{Comparison of $\langle n_{i}\rangle$ and $g^{(2)}_{ij}$ obtained from the numerical exact Monte-Carlo simulation of Eq.~(\ref{eq:rate_eq}) (solid lines) with the ones obtained from the direct solution of the reduced density matrix Eq.~\eqref{eq:reducedrateequ} (dashed lines). Note that the pump in the reduced density matrix is scaled to compensate that $\beta\neq1$.}
 \label{fig:exp_vs_theory_summed}
\end{figure}

\label{app:reduced}
Under the (idealizing) assumption that all spontaneous emission goes into the selected modes i.e.~$\tau^{-1} =0$ and $\beta=1$ a reduced density matrix of the form
$\rho^{n_{h},n_{l}}$ can be derived. To derive the equation of motion for the reduced density matrix we need to consider only the parts of Eq.~\eqref{eq:rate_eq} describing the pump and the photon emission into the cavity
\begin{align}
  \label{eq:reducedrateequ}
  \frac{\diff}{\diff t}\rho^{{\bn}}_N=+&P\left[\rho^{{\bn}}_{N-1}-\rho^{{\bn}}_N\right]+[\dots]
  \\ -&\sum_i g _{i}[N(n_{i}+1)\rho^{\bn}_N-(N+1)n_{i}\rho^{{\bn}-{\bf e_i}}_{N+1}]\nonumber.
\end{align}
Here and in the following equation $[\dots]$ stands for terms that describe
describe the loss of the individual modes as well as intermode transitions.
When no spontaneous emission is lost into non-selected modes the carrier recombination and emission into the cavity modes is faster than any other process so that the term ${\sum_i g_{i}N(n_{i}+1)\rho^{\bn}_N}$ can be substituted by $P\rho^{\bn}_{N-1}$. This means that whenever an emitter is excited by the pump its excitation is immediately emitted into the cavity, thus the emission into the modes can be described directly by $P\rho^{\bn}_{N-1}$ \cite{rice_photon_1994}.
By the same reasoning or by simply shifting the indices (${\rho^{\bn-{\bf e_i}}_{N+1}}$) one can find a substitute for all terms in  Eq.~(\ref{eq:rate_eq}) that correspond to the photon emission. Now we can trace over the emitter subspace ($\sum_N$), resulting in an equation of motion for the reduced density matrix,
\begin{align*}
  \frac{\diff}{\diff t}\rho^{{\bn}}=-&P\rho^{{\bn}}\nonumber
  +\sum_i\frac{P\,g_{i}n_{i}\rho^{\bn-{\bf e_i}}}{\sum_{j}g_{j}(n_{j}+1)-g_{i}}+[\dots].
\end{align*}
As argued above, the reduced density matrix approach works under the assumption of $\beta=1$. In our case $\beta\simeq0.2$, so only a fraction of the pump effectively creates photons in the cavity. To be able to compare the reduced density matrix to the full model the pump power is scaled accordingly. Figure~\ref{fig:exp_vs_theory_summed} shows the results of the Monte-Carlo simulations of the full equation compared to the results obtained from numerical solution of the reduced equation. The deviation of the reduced density matrix approach for small pump rates is not a problem since for low pump rates the full equation can still be solved numerically exactly (as it is done for Fig.~\ref{fig:photon_statistic} A and B). Importantly, the reduced equation reproduces the results in the regime of high pump rates, where the exact solution for the full equation can no longer be obtained.

\section{The role of spontaneous intermode transitions}
\label{app:factorization}

\begin{figure}
  \includegraphics[width=0.95\linewidth]{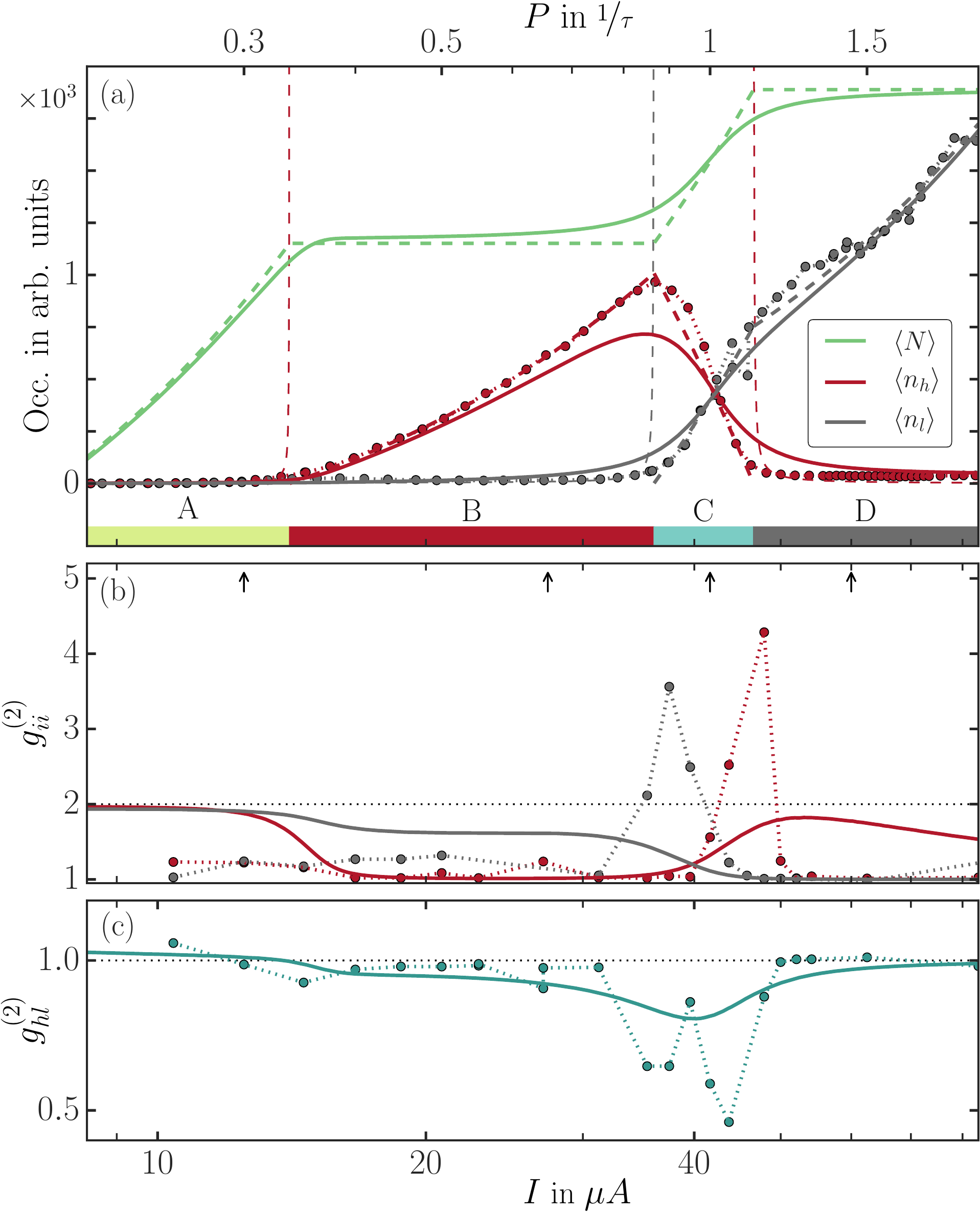}
  \caption{Calculated input-output characteristics and fluctuation for both modes in the presence of spontaneous transitions between the modes, $s=1$. The fluctuation are restricted to values $g^{(2)}_{ii}\le 2$. All parameters are chosen as in Fig.~\ref{fig:exp_vs_theory}.}
  \label{fig:exp_vs_theory_s}
\end{figure}

In this appendix we investigate the impact of the effective parameter $s$, which
quantifies spontaneous intermode transitions.
Figure \ref{fig:exp_vs_theory_s} shows the mode characteristics for the case with full spontaneous transitions between the modes ($s = 1$).
In contrast to Fig.~\ref{fig:exp_vs_theory}, the sharp kinks in the occupations
[panel (a)] and the cross correlation [panel (c)] in phase C are less pronounced.
However, the most significant deviation appears in the photon autocorrelations.
For $s=1$ the computed autocorrelation does not reproduce the experimentally
observed superthermal fluctuations, $g^{(2)}\le 2$.

This numerical observation 
can be backed analytically using the following argument. In phase B and D the
joint distribution $\rho^{{\bn}}_N$ factorizes approximately into two parts, one
describing the non-selected mode $i$ and the other one describing the selected
mode $j$ and the emitters, $\rho^{{\bn}}_N \approx \rho^{n_i}\rho^{n_j}_N$.
This statement is confirmed by the numerical solution of the full master equation.
Such a factorization allows one, to trace out both the selected mode and the
emitters, to obtain an equation of motion for the photon number distribution of
the non-selected mode:
\begin{align}
  \frac{\diff}{\diff t}\rho^{n}&=-\N g[(n+1)\rho^{n}-n\rho^{n-1}]\nonumber\\
      -& \ell[n\rho^{n}-(n+1)\rho^{n+1}]\nonumber   \\
       -&R_{j\to i}\n{j}\big[(n+s)\rho^{n} -(n-1+s)\rho^{n-1}\big]\nonumber\\
 -&R_{i\to j}(\n{j}+s)\big[n\rho^{n}-(n+1)\rho^{n+1}\big].\label{eq:factrate_eq}
\end{align}
Here, $n$ denotes the photon number of the non-selected mode, $\n{j},\N$ the mean
occupations of the selected mode and the emitters respectively,
$R_{j\to i} (R_{i\to j})$ the transition rate from (to) the selected mode $j$ to
(from) the non-selected mode, and $g$, $\ell$ the gain and loss rate of the non
selected mode. If spontaneous transitions are fully included, $s=1$,
Eq.~\eqref{eq:factrate_eq} is solved by a distribution of the form
$\rho^{n}= (1-\alpha)\alpha^{n}$, which always yields $g^{(2)}=2$. This explains
why we do not find superthermal fluctuations for $s=1$.


%

\end{document}